\documentclass[twocolumn,aps,prl,preprintnumbers,superscriptaddress]{revtex4}
\usepackage[latin9]{inputenc}
\usepackage{amsmath}
\usepackage{amssymb}
\usepackage{graphicx}
\usepackage{esint}

\makeatletter
\@ifundefined{textcolor}{}
{%
 \definecolor{BLACK}{gray}{0}
 \definecolor{WHITE}{gray}{1}
 \definecolor{RED}{rgb}{1,0,0}
 \definecolor{GREEN}{rgb}{0,1,0}
 \definecolor{BLUE}{rgb}{0,0,1}
 \definecolor{CYAN}{cmyk}{1,0,0,0}
 \definecolor{MAGENTA}{cmyk}{0,1,0,0}
 \definecolor{YELLOW}{cmyk}{0,0,1,0}
}

\usepackage{ulem}\usepackage{float}

\newcommand{\avg}[1]{\left< #1 \right>} % for average

\makeatother

\begin{document}

\title{Observation of Incipient Charge Nematicity in Ba(Fe$_{1-x}$Co$_{x})_{2}$As$_{2}$}

\author{Y. Gallais}
\email{yann.gallais@univ-paris-diderot.fr}
\affiliation{Laboratoire Matériaux et Phénom$\grave{e}$nes Quantiques
(UMR 7162 CNRS), Université Paris Diderot-Paris 7, Bât. Condorcet,
75205 Paris Cedex 13, France}
\author{R. M. Fernandes}
\affiliation{School of Physics and Astronomy, University of Minnesota,
Minneapolis, 55455, USA}
\author{I. Paul}
\affiliation{Laboratoire Matériaux et Phénom$\grave{e}$nes Quantiques
(UMR 7162 CNRS), Université Paris Diderot-Paris 7, Bât. Condorcet,
75205 Paris Cedex 13, France}
\author{L. Chauvière}
\affiliation{Laboratoire Matériaux et Phénom$\grave{e}$nes Quantiques
(UMR 7162 CNRS), Université Paris Diderot-Paris 7, Bât. Condorcet,
75205 Paris Cedex 13, France}
\author{Y.-X. Yang}
\affiliation{Laboratoire Matériaux et Phénom$\grave{e}$nes Quantiques
(UMR 7162 CNRS), Université Paris Diderot-Paris 7, Bât. Condorcet,
75205 Paris Cedex 13, France}
\author{M.-A. Méasson}
\affiliation{Laboratoire Matériaux et Phénom$\grave{e}$nes Quantiques
(UMR 7162 CNRS), Université Paris Diderot-Paris 7, Bât. Condorcet,
75205 Paris Cedex 13, France}
\author{M. Cazayous}
\affiliation{Laboratoire Matériaux et Phénom$\grave{e}$nes Quantiques
(UMR 7162 CNRS), Université Paris Diderot-Paris 7, Bât. Condorcet,
75205 Paris Cedex 13, France}
\author{A. Sacuto}
\affiliation{Laboratoire Matériaux et Phénom$\grave{e}$nes Quantiques
(UMR 7162 CNRS), Université Paris Diderot-Paris 7, Bât. Condorcet,
75205 Paris Cedex 13, France}
\author{D. Colson}
 \affiliation{CEA-Saclay, IRAMIS, Service de Physique de l'Etat Condensé
(SPEC URA CNRS 2464), F-91191 Gif-sur-Yvette, France}
\author{A. Forget}
\affiliation{CEA-Saclay, IRAMIS, Service de Physique de l'Etat Condensé
(SPEC URA CNRS 2464), F-91191 Gif-sur-Yvette, France}

%\date{\today}
\begin{abstract}
Using electronic Raman spectroscopy, we report direct measurements
of charge nematic fluctuations in the tetragonal phase of \textit{strain-free}
Ba(Fe$_{1-x}$Co$_{x})_{2}$As$_{2}$ single crystals. The strong enhancement of the
Raman response at low temperatures unveils an underlying charge nematic
state that extends to superconducting compositions and which has hitherto remained unnoticed.
Comparison between the extracted charge nematic susceptibility and the elastic modulus allows
us to disentangle the charge contribution to the nematic instability, and to show that charge nematic fluctuations are weakly coupled to the lattice.
\end{abstract}
\maketitle
Electronic analogues of nematic states, in which rotational symmetry
is broken but translational invariance is preserved, have been proposed
in a variety of correlated materials~\cite{Fradkin}, such as quantum
Hall systems~\cite{Lilly}, cuprates~\cite{Kivelson-RMP,Daou},
ruthenates~\cite{Borzi}, heavy fermions~\cite{Okazaki} and, more
recently, iron pnictide superconductors~\cite{Chu,Kasahara}. In
the latter, several experiments~\cite{Chu,Tanatar,Davis,ARPES-Yi,Dusza,Chu-2}
on strained samples have collected strong but indirect evidence that
the tetragonal-to-orthorhombic structural transition is driven not
by the lattice, but by electronic nematicity. However, these measurements
could not disentangle the roles of the spin~\cite{Fang,Xu,Fernandes,Paul},
charge and orbital~\cite{Lee,Devereaux,Phillips,Kontani} degrees
of freedom in the nematic instability.

In Ba(Fe$_{1-x}$Co$_{x})_{2}$As$_{2}$, the structural transition
at $T_{s}$ either precedes or accompanies a magnetic transition at
$T_{N}$, disappearing near the doping concentration with the highest
superconducting transition temperature $T_{c}$ (see the phase diagram
of Fig. \ref{polar}(a)). The nematic/orthorhombic state is characterized
by inequivalent Fe-Fe bond lengths along the in-plane $a$ and $b$
directions ($x$ and $y$ coordinates respectively of the one-Fe unit
cell used throughout, see Fig. \ref{polar}(a)), and by anisotropic
electronic properties~\cite{Chu,Tanatar,ARPES-Yi,Dusza,Fernandes_SUST}.
If this state is indeed a consequence of the condensation of an electronic
nematic order parameter, its fluctuations should be present in the
tetragonal phase and should increase as the temperature is lowered
towards $T_{s}$. Probing these electronic nematic fluctuations directly
is therefore fundamental to unveil the nature of the structural transition,
and to evaluate their possible role in the superconducting pairing
mechanism.

Here, we report electronic Raman scattering measurements of the charge
nematic susceptibility in the tetragonal phase of Ba(Fe$_{1-x}$Co$_{x})_{2}$As$_{2}$
single crystals in which no explicit tetragonal symmetry breaking stress was applied (i..e. strain-free crystals).
We show that charge nematic fluctuations are manifested
in the Raman spectra by a quasi-elastic peak in the $x^{2}-y^{2}$
(B$_{1g}$) symmetry, whose intensity strongly increases in the tetragonal
phase upon approaching $T_{s}$, signaling an incipient charge nematic
order. The extracted static charge nematic susceptibility displays
a sizable enhancement over a wide doping range above the superconducting
dome, suggesting it may play a role in the superconducting mechanism.
Comparison with available shear modulus data indicates that the enhanced
charge nematic susceptibility is weakly coupled to the lattice, 
highlighting the need to incorporate additional degrees of freedom to explain the structural transition.

\begin{figure}
\centering \includegraphics[width=0.98\linewidth]{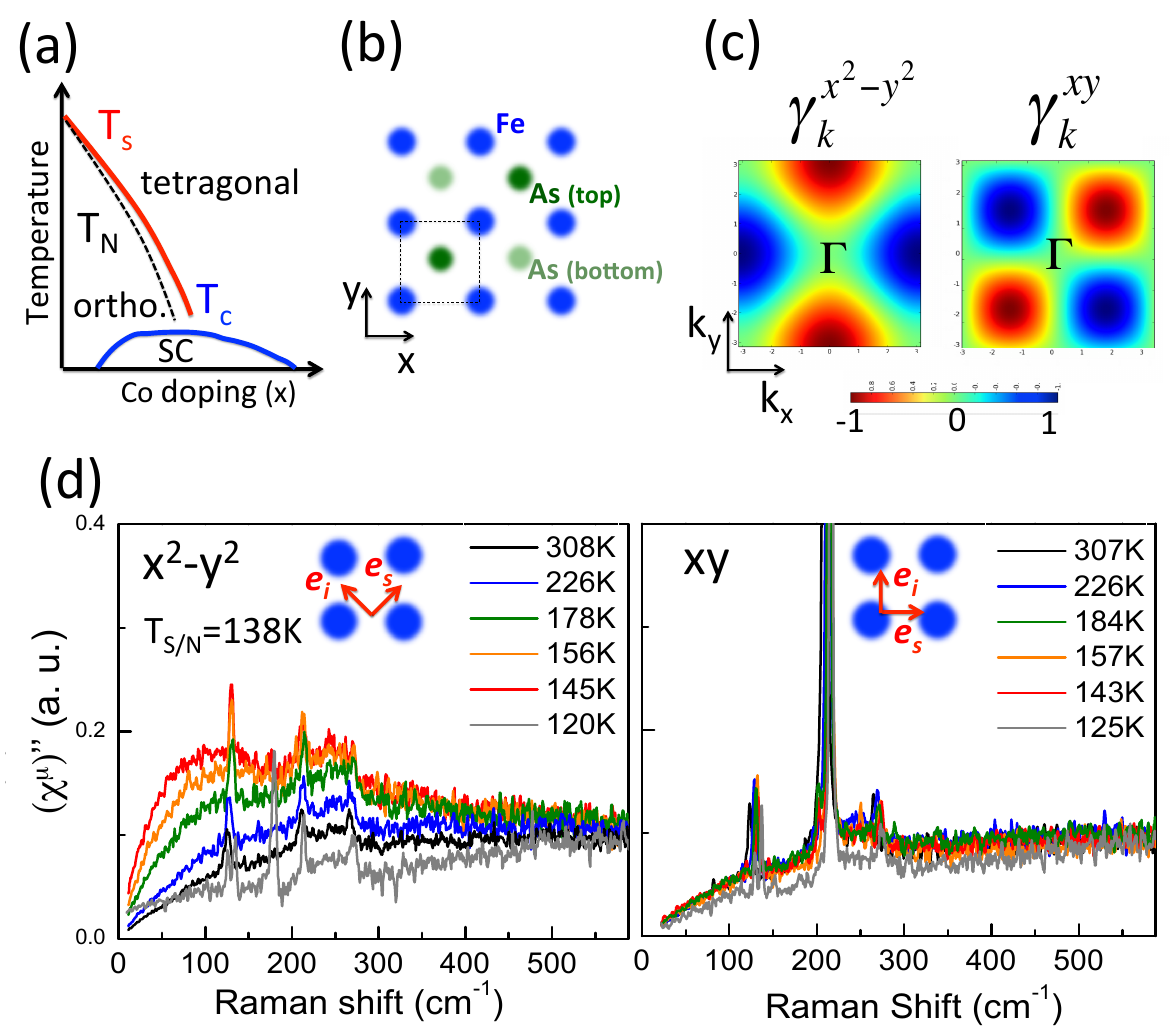} \caption{(a), Sketch of the phase diagram of Ba(Fe$_{1-x}$Co$_{x})_{2}$As$_{2}$. $T_{s}$, $T_{N}$ and $T_{c}$ are the structural, magnetic and superconducting (SC) transition temperatures respectively. (b), Tetragonal FeAs layer, with the $x$ and $y$ axes defined along the Fe-Fe bonds (c), Momentum-space
structure of the form factor $\gamma_{\mathbf{k}}^{\mu}$ for $x^{2}-y^{2}$
and $xy$ symmetries \cite{SI4}. (d), Temperature dependent Raman response
$(\chi^{x^{2}-y^{2}})''$ and $(\chi^{xy})''$ in a strain-free BaFe$_{2}$As$_{2}$
single crystal with $T{}_{s}=138\:$K. The incoming and outgoing photon
polarizations (${\mathbf{e_{I}}}$ ,${\mathbf{e_{S}}}$) used for
each symmetry configuration are depicted in the insets. The sharp
peaks are due to phonon excitations. The electronic Raman continuum
in $x^{2}-y^{2}$ symmetry displays a low frequency quasi-elastic
peak (QEP) that is superimposed on a weaker and broad continuum that
extends to energies above $1000\:\mathrm{cm^{-1}}$ and is essentially
temperature independent in the tetragonal phase (see supplemental Material \cite{SI}). In the orthorhombic
phase, this broad continuum shows a suppression below $500\:\mathrm{cm^{-1}}$
in both symmetries because of the Fermi surface reconstruction induced
by the simultaneous magnetic order \cite{Chauviere}.}
\label{polar} 
\end{figure}

Raman experiments have been carried out using a diode-pumped solid
state laser emitting at 532$\:\mathrm{nm}$ and a triple grating spectrometer
equipped with a nitrogen cooled CCD camera \cite{SI}. 
Single crystals of Ba(Fe$_{1-x}$Co$_{x})_{2}$As$_{2}$ were grown using the self-flux
method. The magnetic and superconducting transition temperatures were
determined by transport measurements performed on crystals from the
same batch \cite{Rullier}. The structural transition temperature
was determined by monitoring phonon anomalies observed when entering
the orthorhombic phase \cite{Chauviere}.

The electronic Raman response, $\chi^{\mu}$, probes the weighted charge correlation
function $\avg{\rho^{\mu}(\omega)\rho^{\mu}(-\omega)}$, where $\rho^{\mu}=\sum_{\mathbf{k}}\gamma_{\mathbf{k}}^{\mu}n_{\mathbf{k}}$
depends on the charge-density operator $n_{\mathbf{k}}$ of the momentum
state ${\bf k}$, and on the form factor $\gamma_{\mathbf{k}}^{\mu}$
whose symmetry $\mu$ is determined by the polarizations $\mathbf{e}_{I}$
and $\mathbf{e}_{S}$ of the incident and scattered photons \cite{Dev-Hackl,SI4}.
To probe the in-plane charge nematic fluctuations, two polarization
configurations can be considered (see inset of Fig. \ref{polar}(d)).
For photons polarized along the diagonals of the Fe-Fe bonds, the
form factor has $x^{2}-y^{2}$ ($B_{1g}$) symmetry, and
is sensitive to nematic order along the Fe-Fe bonds. This is the type
of $C_{4}$ (tetragonal) symmetry-breaking realized in the iron pnictides.

Note that while the charge nematic order parameter 
$\phi_{\mathbf{k}}\propto\gamma_{\mathbf{k}}^{x^{2}-y^{2}}n_{\mathbf{k}}$ changes sign under a $90^{\circ}$ rotation, $\chi^{\mu}$ is proportional
to its square $\phi_{\mathbf{k}}^{2}$ and therefore is $C_{4}$ symmetric.
Thus, unlike previous transport anisotropy measurements \cite{Chu,Tanatar,Chu-2},
we can extract the nematic fluctuations directly from the Raman response
without applying any external symmetry breaking field such as uniaxial
stress. Besides the $x^{2}-y^{2}$ ($B_{1g}$) symmetry, we also investigated the
form factor with $xy$ ($B_{2g}$) symmetry, which is insensitive
to changes that make $x$ and $y$ inequivalent. The behaviors of
these form factors in momentum space are depicted in Fig. \ref{polar}(c). 
\begin{figure}[t]
\centering \includegraphics[clip,width=0.99\linewidth]{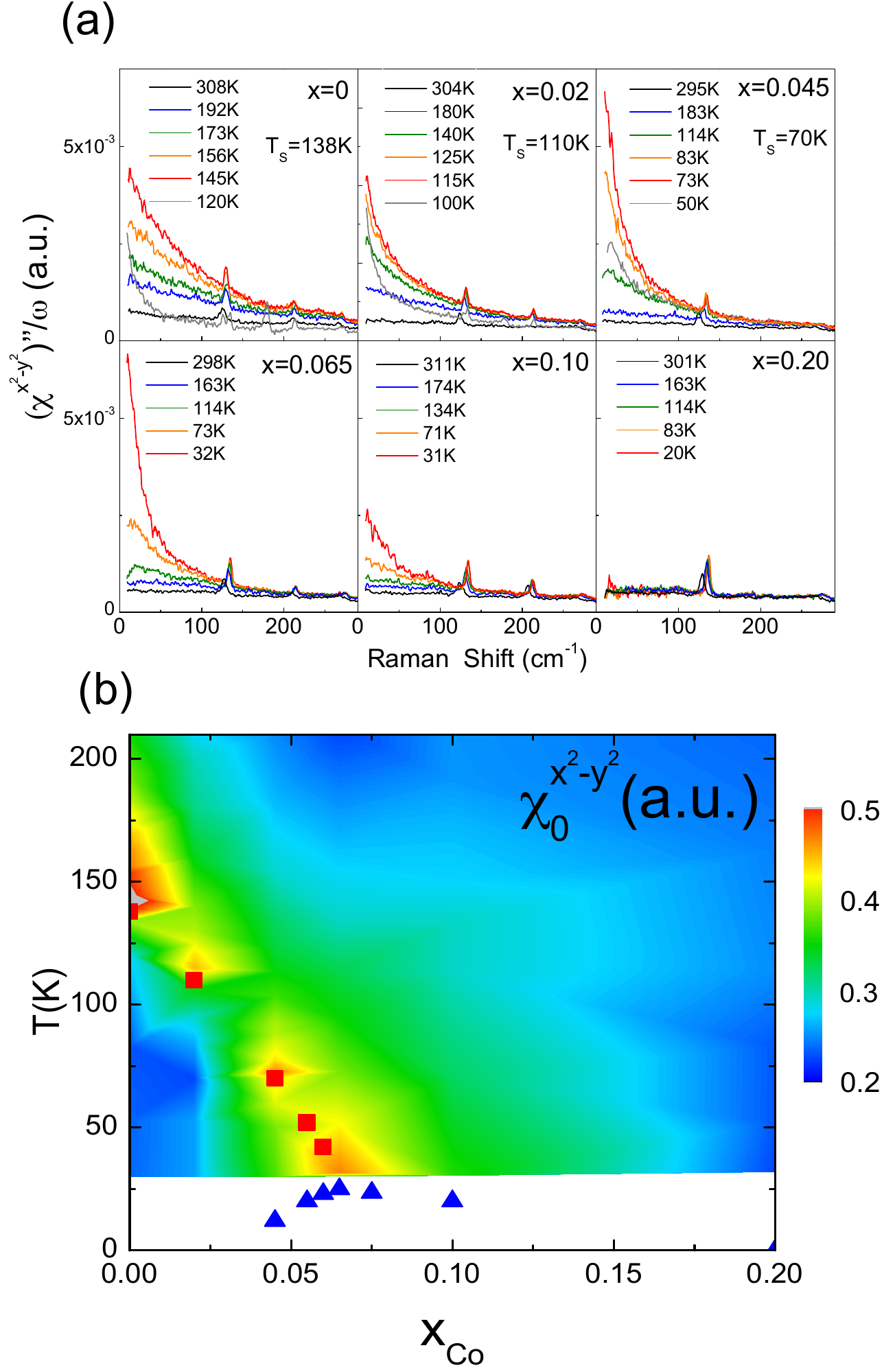} \caption{(a), Temperature dependent Raman conductivity $(\chi^{x^{2}-y^{2}})''/\omega$ for $x=0$ (parent), $x=0.02$ (strongly underdoped), $x=0.045$ (underdoped),
$x=0.065$ (optimally doped), $x=0.10$ (overdoped), and $x=0.20$
(strongly overdoped). The structural transition temperature is indicated
for the three underdoped compositions. The $x=0.065$ composition
corresponds to optimal superconducting transition temperature ($T_{c}=24.5\: K$)
where no structural transition was detected. (b), Evolution
of the static charge nematic susceptibility, $\chi_{0}^{x^{2}-y^{2}}$,
as a function of temperature and doping. The structural transition
temperature $T_{s}$ and the superconducting transition temperature
$T_{c}$ are indicated in red squares and blue triangles respectively.}
\label{conductivity} 
\end{figure}
\par
Because of the fluctuation-dissipation theorem, the dynamic charge nematic
fluctuations should be manifested in the imaginary part of the Raman
response function $\left(\chi^{\mu}\right)''$ in the appropriate
symmetry $\mu$, namely, the $x^{2}-y^{2}$ ($B_{1g}$) symmetry \cite{Tassini,Zeyher}.
This is illustrated in Fig. \ref{polar}(d) for a strain-free, single
crystal of the parent compound BaFe$_{2}$As$_{2}$, where $\left(\chi^{\mu}\right)''$
is plotted as function of frequency for different temperatures and
for the two symmetries described above. While the response in the
$xy$ symmetry is essentially temperature independent above $T_{s}=138\:$K,
the $x^{2}-y^{2}$ response displays a considerable build-up of intensity
below $500\:\mathrm{cm^{-1}}$ upon approaching $T_{s}$, with a subsequent
collapse in the nematic/orthorhombic phase. The temperature dependence
and the distinctive $x^{2}-y^{2}$ symmetry of this low frequency
quasi-elastic peak (QEP) clearly links it to dynamic charge nematic
fluctuations corresponding to an orientational order along the Fe-Fe
bonds. 
While the spectral line shape of the QEP is linked to the relaxational
dynamics of the nematic fluctuations \cite{SI2}, we choose here to
concentrate on a more transparent quantity: the static charge nematic
susceptibility. Indeed the strong increase of the QEP intensity is
associated with an enhanced static charge nematic susceptibility,
$\chi_{0}^{x^{2}-y^{2}}$, via the Kramers-Kronig relation: 
\begin{equation}
\chi_{0}^{x^{2}-y^{2}}=\frac{2}{\pi}\int_{0}^{\infty}d\omega(\chi'')^{x^{2}-y^{2}}(\omega)/\omega\label{KK}
\end{equation}
The relevant quantity governing the static nematic susceptibility
is thus the Raman conductivity $\chi''/\omega$, highlighting the
importance of the low frequency part of $\chi''$ in determining $\chi_{0}^{x^{2}-y^{2}}$.
The temperature dependence of $\chi''/\omega$, where the QEP is now
centered at zero frequency, is shown in Fig. \ref{conductivity}(a)
for six different Co concentrations of Ba(Fe$_{1-x}$Co$_{x})_{2}$As$_{2}$,
spanning the phase diagram from the parent $x=0$ composition ($T_{s}=138\:$K
and $T_{c}=0$) up to the strongly overdoped $x=0.20$ composition
($T_{s}=T_{c}=0$). For $x\leq0.045$, the QEP displays a systematic
enhancement as temperature is lowered towards $T_{s}$ before collapsing
in the symmetry broken phase. The enhancement of the QEP extends
down to $T_{c}$ for $x=0.065$ where the superconducting transition
temperature is optimal and no structural transition is detected.
For this particular composition the QEP was found to disappear quickly
upon entering the superconducting state indicating a suppression of
nematic fluctuations in the superconducting state (not shown).
Above optimal composition, the enhancement of the QEP is strongly reduced
but remains sizable even for $x=0.10$, before disappearing for $x=0.20$.
\begin{figure}
\centering \includegraphics[clip,width=0.99\linewidth]{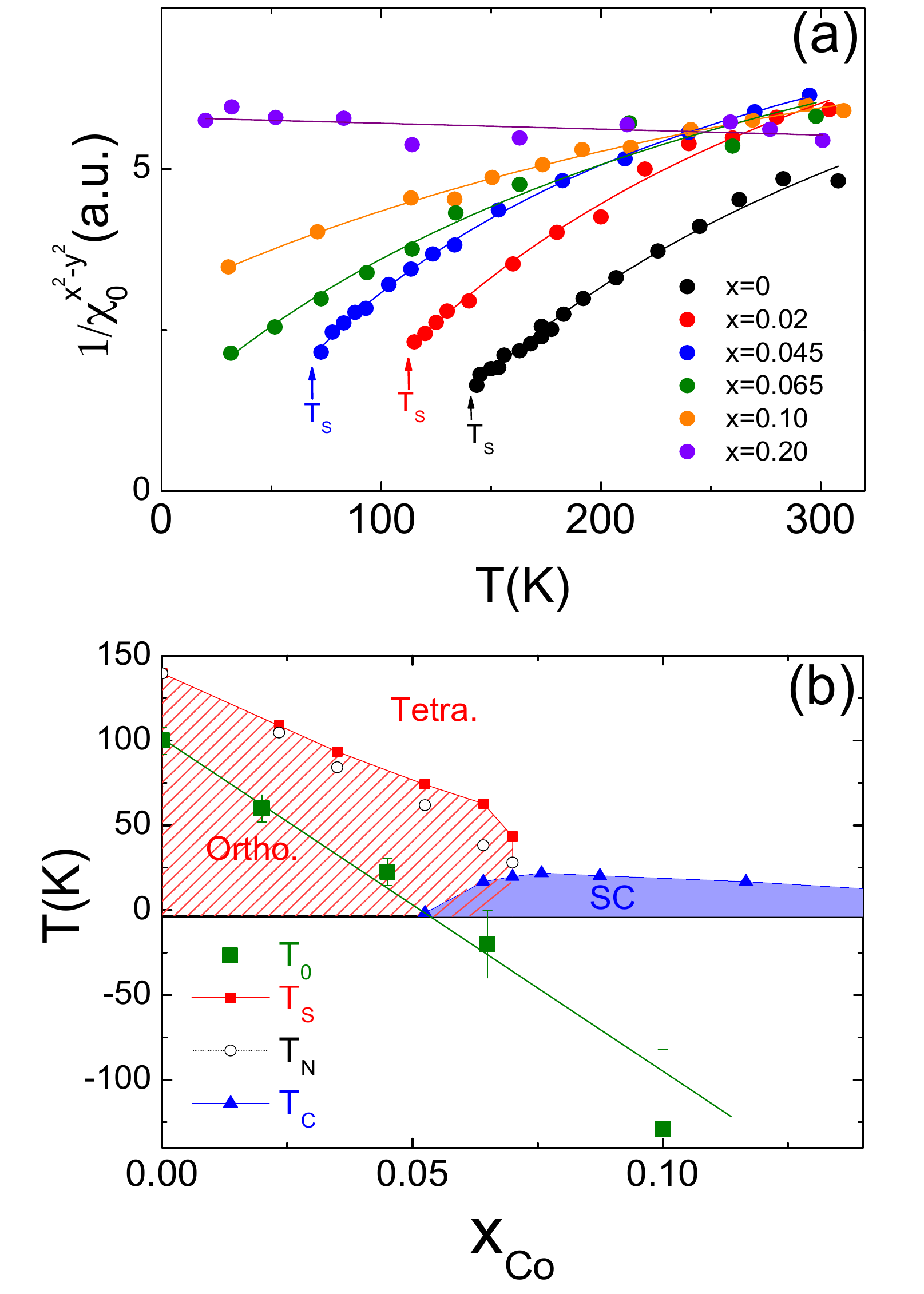} \caption{(a), Temperature dependence of the inverse nematic charge susceptibility,
$(\chi_{0}^{x^{2}-y^{2}})^{-1}$, in the tetragonal phase ($T>T_{s}$)
as a function of Co composition. The lines are Curie-Weiss fits for
each composition (see text). (b), $(x,T)$ phase diagram showing
the orthorhombic (Ortho) and superconducting (SC) phases. The mean-field
transition temperature extracted from the Curie-Weiss fit, $T_{0}$,
is shown in green square (the green line is a linear fit of its doping
dependence). The corresponding structural transition temperature $T_{s}$
(red squares), magnetic transition temperature $T_{N}$ (white circles),
and superconducting transition temperature $T_{c}$ (blue triangles)
are also indicated \cite{Rullier}.}
\label{CW} 
\end{figure}
The static charge nematic susceptibility $\chi_{0}^{x^{2}-y^{2}}$
was extracted using equation (\ref{KK}) via a partial integration of
the Raman conductivity up to $500\:\mathrm{cm^{-1}}$, since above
this frequency the spectra are temperature independent in the tetragonal
phase. To perform the integration, we used a Lorentzian relaxational
form to extrapolate the Raman conductivity spectra from the lowest
frequency experimentally accessible, $9\:\mathrm{cm^{-1}}$, down
to zero \cite{SI}. The doping and temperature dependence of $\chi_{0}^{x^{2}-y^{2}}$
are summarized in the phase diagram of Fig. \ref{conductivity}(b).
The maximum of the static charge nematic susceptibility closely tracks
the structural transition temperature in the underdoped region, vanishing
near optimal doping. This temperature and doping dependence is qualitatively
consistent with previous anisotropic transport data of strained crystals
\cite{Chu,Tanatar,Chu-2}. However, resistivity anisotropy is only
an indirect probe of the nematic order parameter since it cannot disentangle
the various possible sources of electronic nematicity.
\par

To perform a more quantitative analysis, in Fig. \ref{CW}(a) we plot
the inverse susceptibility as a function of temperature in the tetragonal
phase ($T>T_{s}$) and for the six Co compositions. 
The softening of the inverse susceptibility,
is seen for all compositions up to $x=0.10$, being absent only for
the strongly overdoped, non-superconducting, $x=0.20$ composition.
For all other compositions the inverse susceptibility above $T_{s}$
can be well described over a large temperature range, spanning at
least 150$\:$K, by a simple Curie-Weiss law of the form: 
\begin{equation}
\left(\chi_{0}^{x^{2}-y^{2}}\right)^{-1}\left(T\right)=\left(A+\frac{C}{T-T_{0}}\right)^{-1}
\label{suscept}
\end{equation}
 where $A$ and $C$ are constants and
$T_{0}$ is the charge nematic transition temperature. The resulting fits for the inverse susceptibility are
shown in Fig. \ref{CW}(a). They unveil an incipient charge nematic
instability at $T_{0}$ over a wide doping range, which includes the
superconducting dome, in the phase diagram of Ba(Fe$_{1-x}$Co$_{x}$)$_{2}$As$_{2}$.
\par
The extracted $T_{0}$ follows the trend of the thermodynamic structural
transition temperature $T_{s}$, decreasing with doping and vanishing
near $x\sim0.06$. However, $T_{0}$ is significantly smaller than
$T_{s}$, by about $50\:$K, across the entire phase diagram (see
Fig. \ref{CW}(b)). This fact, in conjunction with the observation of
the build-up of the charge nematic fluctuations over a large temperature
range, allow us to conclude that the incipient charge nematicity is
not a mere consequence of the softening of the lattice orthorhombicity
via a static linear coupling. More importantly, since the Curie-Weiss
expression (\ref{suscept}) with $T_{0}$ significantly smaller than
$T_{s}$ describes very well the data up to a few Kelvin above $T_{s}$,
it implies that the incipient charge nematicity is, in fact, weakly
coupled to the lattice. Because the tetragonal symmetry-breaking has
to occur at the same temperature in both elastic and charge degrees
of freedom, our analysis thus suggests the presence of another nematic
degree of freedom which drives the structural transition at $T{}_{s}$.

\begin{figure}
\includegraphics[width=0.9\columnwidth]{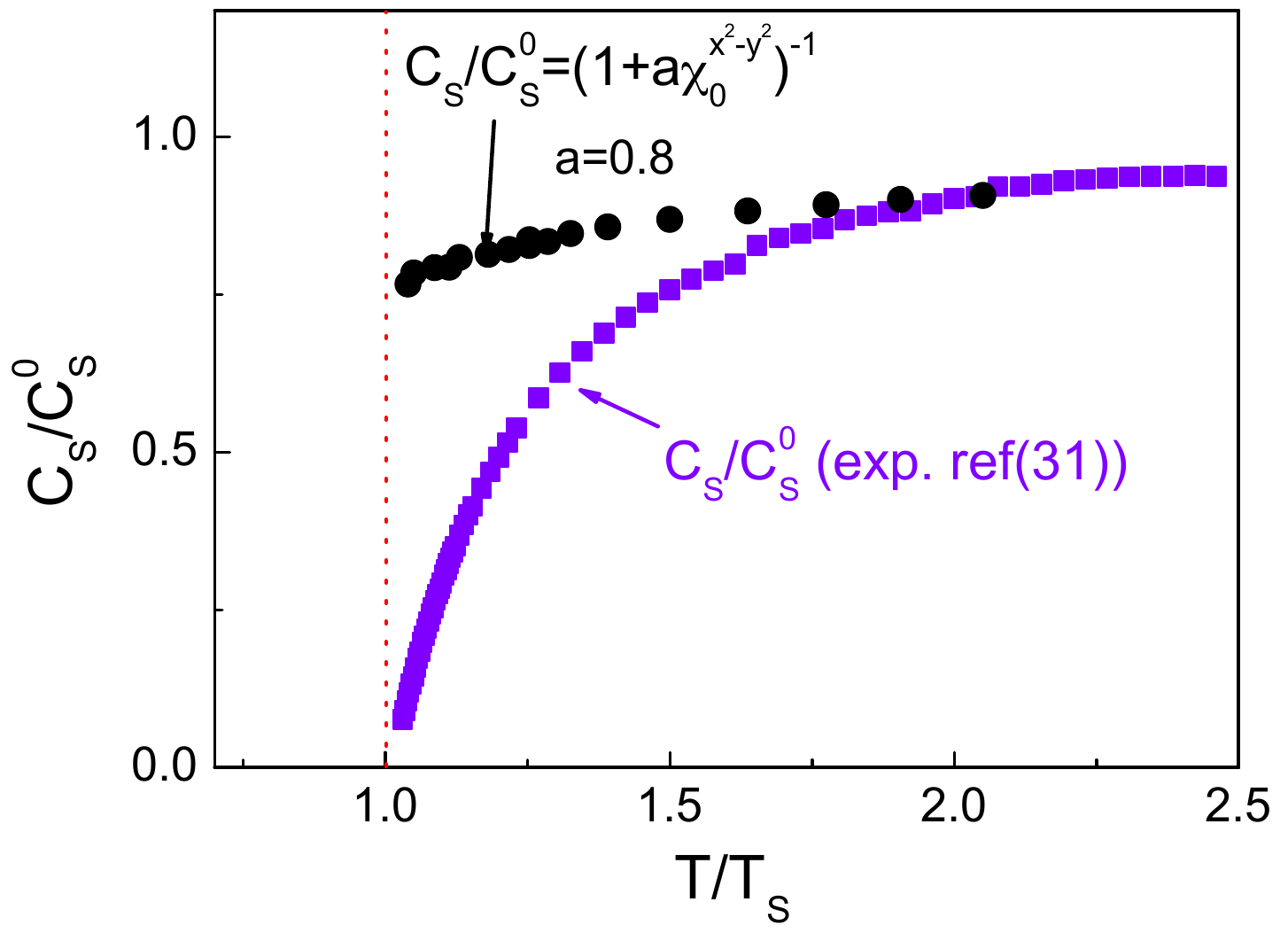} \caption{Temperature dependence of the experimental shear modulus (data of
Ref. \cite{Fernandes-shear}, in purple), together with the expected
temperature dependence of the shear modulus due to the coupling between
the charge nematic and orthorhombic order parameters using Eq.
(\ref{coupling}) (in black). The bare shear modulus was assumed to
be temperature independent and the only adjustable parameter, $a=\frac{\lambda^{2}}{C_{S}^{0}}$
(see text) was chosen to fit the shear modulus data at high temperatures
($a=0.8$). The temperature scale was normalized using the measured
structural transition temperatures (T$_S$=130K in Ref.  \cite{Fernandes-shear}).}
\label{shear} 
\end{figure}

We can also draw the same conclusions simply by comparing the $\chi_{0}^{x^{2}-y^{2}}$
data directly with the shear modulus $C_{s}\equiv C_{11}-C_{12}$,
which measures the orthorhombic lattice stiffness \cite{Fernandes-shear,Yoshizawa}.
Since, by symmetry, the order parameters associated with $C_{s}$
and $\chi_{0}^{x^{2}-y^{2}}$ are linearly coupled, we obtain 
\begin{equation}
\frac{C_{S}}{C_{S}^{0}}=\left[1+\left(\frac{\lambda^{2}}{C_{S}^{0}}\right)\chi_{0}^{x^{2}-y^{2}}\right]^{-1},\label{coupling}
\end{equation}
provided charge nematicity is the only soft mode present \cite{Fernandes-shear,Paul}.
Here, $\lambda$ is the linear coupling constant and $C_{s}^{0}$
is the high-temperature shear modulus. In Fig. \ref{shear}, we test
the validity of the above relation for the parent compound BaFe$_{2}$As$_{2}$
by comparing the $C_{s}$ values inferred from our $\chi_{0}^{x^{2}-y^{2}}$
data via equation (\ref{coupling}) (black) with the experimental $C_{s}$
data of Ref. \cite{Fernandes-shear} (purple). The discrepancy between
the two confirms our inference above that charge nematicity
is not the only soft mode, suggesting the presence of an additional electronic
nematic degree of freedom. The precise nature of this additional degree
of freedom cannot be ascertained from our Raman study. It is possible
that spin fluctuations drive the softening of $C_{s}$ \cite{Fang,Xu,Fernandes,Paul}
and $\chi_{0}^{x^{2}-y^{2}}$ via spin-lattice and spin-charge couplings,
respectively. Alternatively, it has also been proposed that the structural
transition is driven by orbital ordering between $xz$ and $yz$
Fe 3d orbitals \cite{Lee,Devereaux,Phillips,Kontani}. Although the charge
fluctuations measured here do not necessarily come only from fluctuations
of the relative charge $n_{xz}-n_{yz}$ between these two orbitals,
it follows from the orbital content of the Fermi surface of the iron
pnictides that these orbital fluctuations should give a major contribution
to $\chi_{0}^{x^{2}-y^{2}}$ if orbital order is the driving instability
(\cite{Valenzuela}, see also Supplemental Material \cite{SI3}).

In conclusion, we presented electronic Raman spectroscopy study of
Ba(Fe$_{1-x}$Co$_{x})_{2}$As$_{2}$ single crystals in the tetragonal
phase, where the $C_{4}$ symmetry is intact. Our analysis of the
temperature dependence of the enhanced charge nematic susceptibility,
and its comparison to the shear modulus data, indicate that although
these fluctuations contribute to promote the breaking of the tetragonal
symmetry, they are not the only driving mechanism behind it. The persistence
of these fluctuations above the entire superconducting dome raises
the question of whether they play a role in the pairing mechanism
\cite{Fernandes_SUST,Yamase2013}. Our results are reminiscent
of earlier Raman studies indicating fingerprints of fluctuating charge
density wave order in cuprates \cite{Tassini,Blumberg}. We note however
that in contrast to the stripe or checkerboard orders observed in
cuprates, the fluctuating nematic order observed here does not break
any lattice translational symmetry. Besides shedding light on the
nature of the nematic state of the pnictides, our approach provides
a novel route to investigate electronic nematicity in other strongly
correlated systems where this type of state has been proposed.

We thank F. Rullier-Albenque for providing us with transport data.
We acknowledge fruitful discussions with A. Chubukov, V. Keppens,
D. Mandrus and J. Schmalian. Y.G., L.C., Y.X. Y., M.C, M.-A.M., A.S.,
D.C. and A.F. acknowledge support from Agence Nationale de la Recherche
through Grant PNICTIDES.

\newpage

\textbf{\large{Supplemental Material}}

\section{Raman experiments}

Raman experiments have been carried out using a diode-pumped solid
state laser emitting at 532$\:\mathrm{nm}$ and a triple grating spectrometer
equipped with a nitrogen cooled CCD camera. In order to extract the
imaginary part of the Raman response function, the raw spectra were
corrected for the Bose factor and the instrumental spectral response.
All temperatures were corrected for the estimated laser heating.
 It was first estimated by comparing the power and temperature dependences of the phonon frequencies. This
estimate was then cross-checked by monitoring the onset of Rayleigh
scattering by orthorhombic structural domains across the structural
transition temperature as a function of laser power. Both methods
yielded an estimated heating of 1$\:$K $\pm$ 0.2 per mW of incident
power. In order to extract the imaginary part of the Raman response
function, the raw spectra were corrected for the Bose factor and the
instrumental spectral response.

\section{Extraction of the static charge nematic susceptibility from Raman
scattering measurements}

The experimentally measured Raman intensity in the symmetry $\mu$,
$I^{\mu}(\omega)$, is proportional to the weighted charge correlation
function $S^{\mu}(\omega$).

\begin{equation}
I^{\mu}(\omega)\propto S^{\mu}(\omega)=\avg{\rho^{\mu}(\omega)\rho^{\mu}(-\omega)}\label{correlation}
\end{equation}
 The correlation function $S^{\mu}$ is in turn directly linked to
the imaginary part of the Raman response $(\chi^{\mu})''$ via the
fluctuation dissipation theorem:

\begin{equation}
S^{\mu}(\omega)=-\frac{\hbar}{\pi}(1+n(\omega,T))(\chi^{\mu})''(\omega)\label{bose}
\end{equation}
 where $n(\omega,T)$ is the Bose-Einstein distribution function.
The electronic Raman response function $\chi^{\mu}$ is given by:

\begin{equation}
\chi^{\mu}(\omega)=\frac{i}{\hbar}\int_{0}^{\infty}dte^{i\omega t}\left<[\rho^{\mu}(t),\rho^{\mu}(0)]\right>\label{response}
\end{equation}
 where the operator $\rho^{\mu}$ has the form 
\begin{equation}
\rho^{\mu}=\sum_{\mathbf{k}}\gamma_{\mathbf{k}}^{\mu}n_{\mathbf{k}}\label{density}
\end{equation}
 in terms of the charge density operator $n_{\mathbf{k}}$ in the
momentum state $\mathbf{k}$, and a form factor, also called the Raman
vertex, $\gamma_{\mathbf{k}}^{\mu}({\bf {e_{i},e_{s}})}$, whose symmetry
index $\mu$ depends on the polarizations of the incident and scattered
lights ${\bf{e_{i}}}$ and ${\bf{e_{s}}}$ respectively.~\cite{Dev-Hackl} Since
the photon wave vector is several orders of magnitude smaller than
the typical Brillouin zone size, there is negligible momentum transfer
in the electron-photon scatterings, and therefore Raman spectroscopy
probes the system uniformly.

The $x^{2}-y^{2}$ or $B_{1g}$ symmetry can be selected by choosing
crossed incoming and outgoing photon polarizations at 45 degrees with
respect to the Fe-Fe bonds. Here the notation $B_{1g}$ refers to
the one Fe unit cell whose axes are along the Fe-Fe bonds. Similarly
the $xy$ or $B_{2g}$ symmetry can be selected by choosing incoming
and outgoing photon polarizations along the Fe-Fe bonds. Note that
in terms of the full lattice unit cell (or 4 Fe unit cell), which
has its axes at 45 degrees to the Fe-Fe bonds and is more commonly
used in the Raman literature, the $B_{1g}$ ($B_{2g}$) symmetry discussed
here corresponds to the $B_{2g}$ ($B_{1g}$) symmetry.

The momentum space structure of $\gamma_{\mathbf{k}}^{\mu}$ is constrained
by symmetry. For example in the case of $x^{2}-y^{2}$ symmetry, $\gamma_{\mathbf{k}}^{\mu}$
must change sign under mirror symmetry with respect to the direction
at 45 degrees of the $x$ and $y$ axis. Using the effective mass
approximation for a tight binding model with nearest neighbour hopping
integrals only we have $\gamma_{\mathbf{k}}^{x^{2}-y^{2}}=\cos k_{x}-\cos k_{y}$
and $\gamma_{\mathbf{k}}^{xy}=\sin k_{x}\sin k_{y}$~\cite{Dev-Hackl,Valenzuela}.
The k-space structure of these form factors are shown in Fig. 1 of
the main manuscript.

While we only have access to the imaginary part of the symmetry dependent
response as a function of frequency, we can extract the corresponding
static susceptibility, $\chi_{0}^{\mu}$, using the Kramers-Kronig
relation linking the real and the imaginary parts of the Raman response
function: 
\begin{equation}
\chi'(\omega)=\frac{1}{\pi}\int_{-\infty}^{\infty}d\omega'\frac{\chi''(\omega')}{\omega'-\omega}
\end{equation}
 Taking $\omega=0$ and using the fact that $\chi''$ is an odd function
of $\omega$, we obtain the following expression for the symmetry
dependent static susceptibility: 
\begin{equation}
\chi_{0}^{\mu}=\frac{2}{\pi}\int_{0}^{\infty}d\omega'(\chi^{\mu})''(\omega')/\omega'\label{KK}
\end{equation}
The expression above shows that the relevant quantity controlling
the static susceptibility is not the Raman response $\chi''$ but
the Raman ''conductivity`` $\chi''/\omega$
which is dominated by its low frequency behavior.

\begin{figure}
\centering \includegraphics[clip,width=0.95\linewidth]{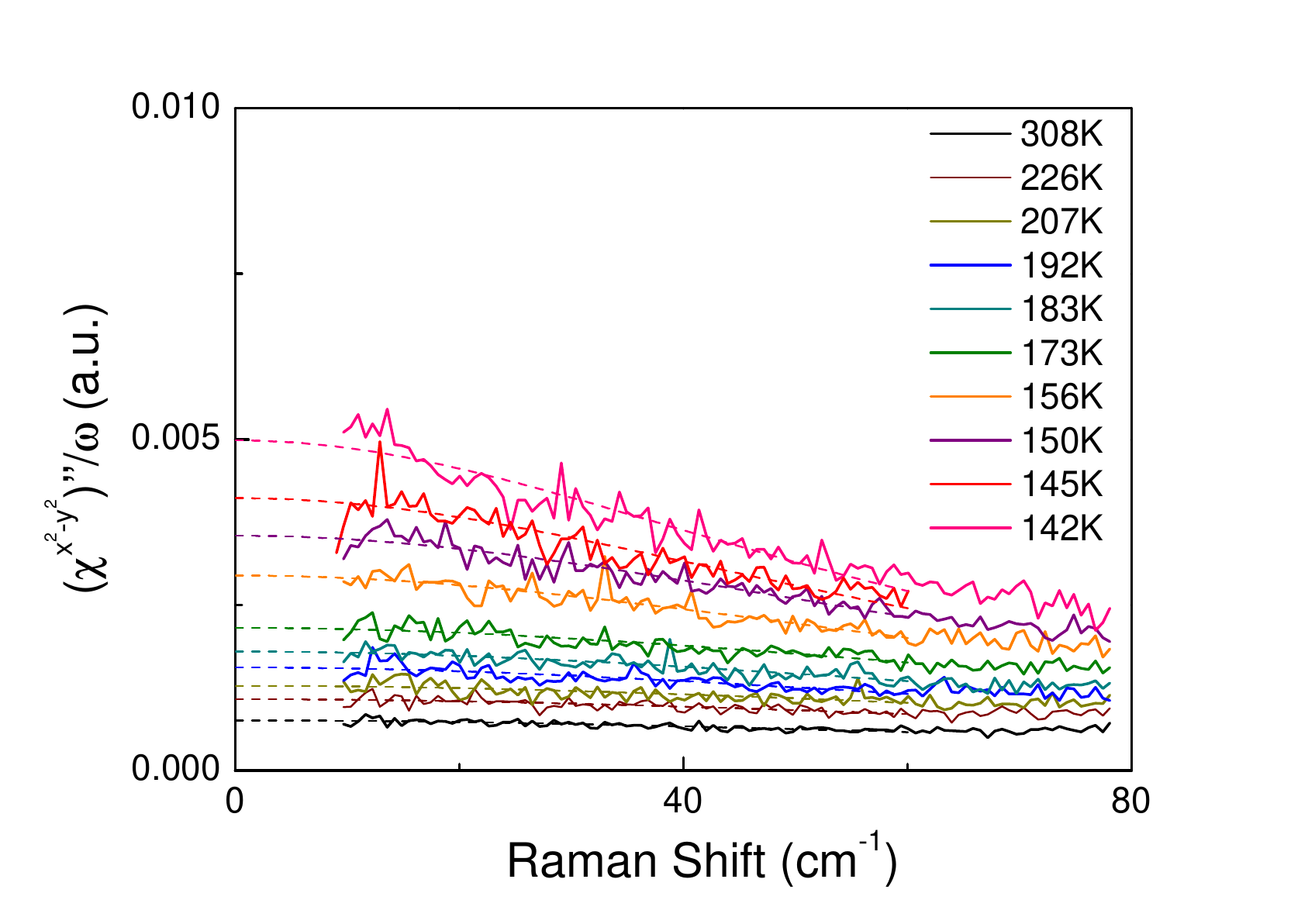}
\caption{Extrapolation of $(\chi^{x^{2}-y^{2}})''/\omega$ for $x=0$ (parent
compound BaFe$_{2}$As$_{2}$) and for 10 different temperatures in
the tetragonal phase. The extrapolation was performed by fitting with
a Lorentzian relaxational form the frequency window 9-25 $\:\mathrm{cm^{-1}}$
(see text).}
\label{conductivity} 
\end{figure}

In terms of the generalized momentum-dependent charge nematic response
function $\chi^{\mu}({\bf {q},\omega)}$, the above procedure is equivalent
to taking the limit ${\bf {q}\rightarrow0}$ first and then $\omega\rightarrow0$.
On the other hand the static nematic susceptibility, which diverges
at a second order phase transition, is defined by the opposite order
of limits where $\omega\rightarrow0$ first and then ${\bf {q}\rightarrow0}$.
It is well-known that the order of limits is crucial for conserved
quantities, for which the former way of taking limits give zero while
the latter gives a finite value in thermodynamically stable phases.
It is important to note that the nematic charge operator defined in
Eq.~(\ref{density}) is not a conserved quantity, and therefore the
two limits can be interchanged.

Experimentally only the response in $x^{2}-y^{2}$ symmetry, which
measures charge nematic fluctuations, shows a significant build-up
at low frequency or quasi-elastic peak (QEP) upon cooling. The QEP
appears as a peak centered at zero-frequency in the raw intensity
data, $I$, while it is pushed to finite frequency when the quantity
$\chi''$ is plotted as in the main manuscript (see Eq. (\ref{bose})).
Since we have only access to a finite frequency range, the integral
in Eq. (\ref{KK}) can only be performed up to a finite frequency cut-off.
Experimentally the QEP above T$_{s}$ is superimposed on a weaker
and broad electronic continuum that extends up to energies above 1000$\:\mathrm{cm^{-1}}$.
This continuum shows a reconstruction in the SDW state \cite{chauviere} in both $x^{2}-y^{2}$ and $xy$ symmetries
but is essentially temperature independent in the tetragonal phase. This suppression in the SDW state demonstrates a sizeable electronic Raman response in both symmetries, but the presence of nematic fluctuations in the $x^{2}-y^{2}$ symmetry only.
\par
In the tetragonal phase $\chi''$ is to within our experimental accuracy temperature
independent above 500$\:\mathrm{cm^{-1}}$ in both symmetries.
One can therefore reliably extract the temperature dependent charge
nematic susceptibility by restricting the integral to energies lower
than 500 cm$^{-1}$ where the response in temperature dependent and
dominated by the QEP. On the low energy side, the Raman measurements
were performed down to 9 $\:\mathrm{cm^{-1}}$. In order
to perform the integration down to zero frequency, the Raman conductivity,
$(\chi^{x^{2}-y^{2}})''(\omega')/\omega'$ was extrapolated assuming
a Lorentzian relaxational form for the Raman conductivity at low frequency:

\begin{equation}
(\chi^{x^{2}-y^{2}})''(\omega)/\omega\sim\frac{\Gamma}{\Gamma^{2}+\omega^{2}}
\end{equation}
 where $\Gamma$ is a static relaxation rate. The extrapolations are
shown in Fig. \ref{conductivity}. The resulting static nematic susceptibility
is shown as a function of temperature and doping in Fig. 2 of the
manuscript.

\section{Modeling of the quasi-elastic peak using a Lorentzian relaxational
form}

The data in $x^{2}-y^{2}$ symmetry channel can be modeled
using a temperature independent background of the same shape as the
one seen in the $xy$ symmetry channel plus a quasi-elastic peak (QEP)
which has the functional form of a Lorentzian relaxation with a scattering
rate $\Gamma$: 

\begin{equation}
(\chi^{x^{2}-y^{2}})''\sim\frac{\omega\Gamma}{\Gamma^{2}+\omega^{2}}
\end{equation}

\begin{figure*}
\centering \includegraphics[clip,width=0.75\linewidth]{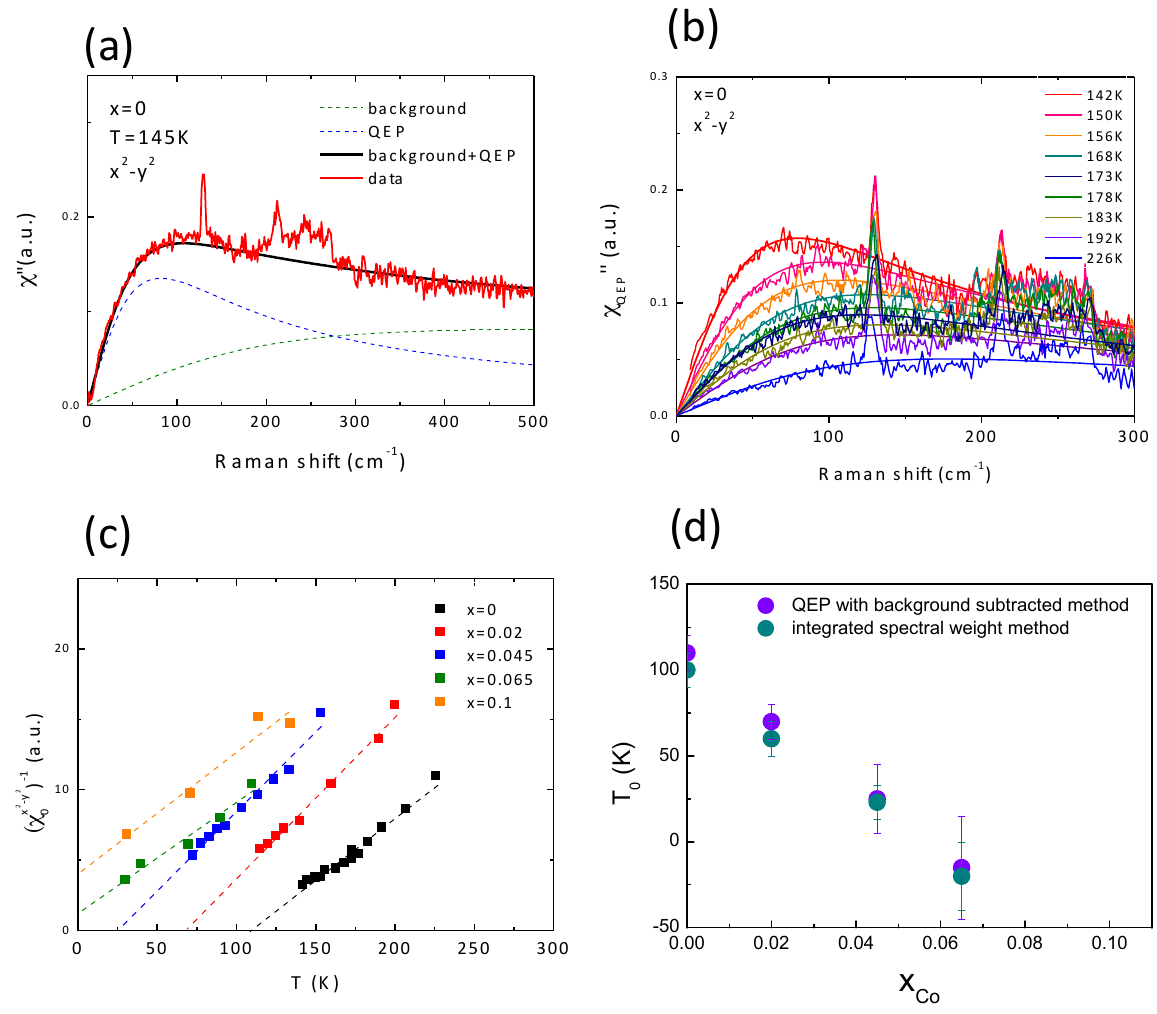}
\caption{(a): decomposition of the Raman spectrum in $x^{2}-y^{2}$
symmetry using a broad continuum and a Lorentzian QEP. The broad continuum
was taken so as to fit the $xy$ continuum in the tetragonal phase.
(b): fits of the background substracted QEP as a function
of temperature for x=0 using the Lorentzian relaxational form described
in the text. (c): extracted inverse nematic susceptibility
by integrating only the extracted diverging QEP part of the Raman
conductivity for $x=0$, $x=0.02$, $x=0.045$, $x=0.065$ and $x=0.1$. The inverse
susceptibility shows linear Curie-Weiss-like behavior. $T_{0}$ corresponds
to the temperature at which the inverse susceptibility extrapolates
to zero. (d): x dependence of T$_0$ extracted from this analysis (in purple) and from the 
analysis described in the main text (in green). The T$_0$ values agrees with 10 percent.}
\label{QEP} 
\end{figure*}

This analysis is shown in figure \ref{QEP}(a),(b) below for x=0. The data in the tetragonal phase are well reproduced by this modeling for all temperatures. One advantage of this analysis
is that it allows one to extract the spectral weight of the QEP only:
the background-free diverging part of the charge nematic susceptibility $\chi_0^{x^{2}-y^{2}}$.
One drawback however, is that the nematic susceptibility can only
be reliably extracted when the QEP intensity is sizable, i.e. close to $T_{s}$, in our case between $T_{s}$ and $T_{s}$ +100K approximately.
Outside this temperature range the fits and the resulting analysis
are too dependent on the exact lineshape of the background that is
subtracted.

The inverse susceptibility extracted with this method is
shown in figure \ref{QEP}(c). It follows reasonably well
a linear (i.e. Curie-Weiss) behavior at least for x=0, x=0.02, x=0.045
and x=0.065 compositions: $(\chi_0^{x^{2}-y^{2}})^{-1}\sim T-T_0$. This is fully consistent with the analysis
reported in the main text where the total susceptibility (background
plus QEP) was fitted with a constant plus a Curie Weiss term (see Eq. (2) of the main text). Besides, as shown in figure \ref{QEP}(d)
the extracted temperature $T_{0}$ from the linear fits agrees within error bars with the one extracted from the analysis performed in the main text.

\section{Polarization resolved spectra for $x=0.02$}

We show in fig. \ref{polar} the full polarization resolved
Raman spectra for x=0.02 at T=115K. The four polarization configurations include
the two crossed polarization configurations introduced in the manuscript
(B$_{1g}$ or $x^{2}-y^{2}$ symmetry, and $xy$
or B$_{2g}$ symmetry) along with the two parallel polarizations configurations
which probe the A$_{1g}$+B$_{1g}$ and the A$_{1g}$+B$_{2g}$ symmetries of the tetragonal lattice structure.
All symmetries show a sizable continuum that extends to energies above
500$\:\mathrm{cm^{-1}}$ but the temperature dependent low energy
spectral weight due to the quasi-elastic peak is only seen in 
the two configurations (red and orange) which probe the B$_{1g}$ or $x^{2}-y^{2}$
symmetry. 

\begin{figure}
\centering \includegraphics[clip,width=0.95\linewidth]{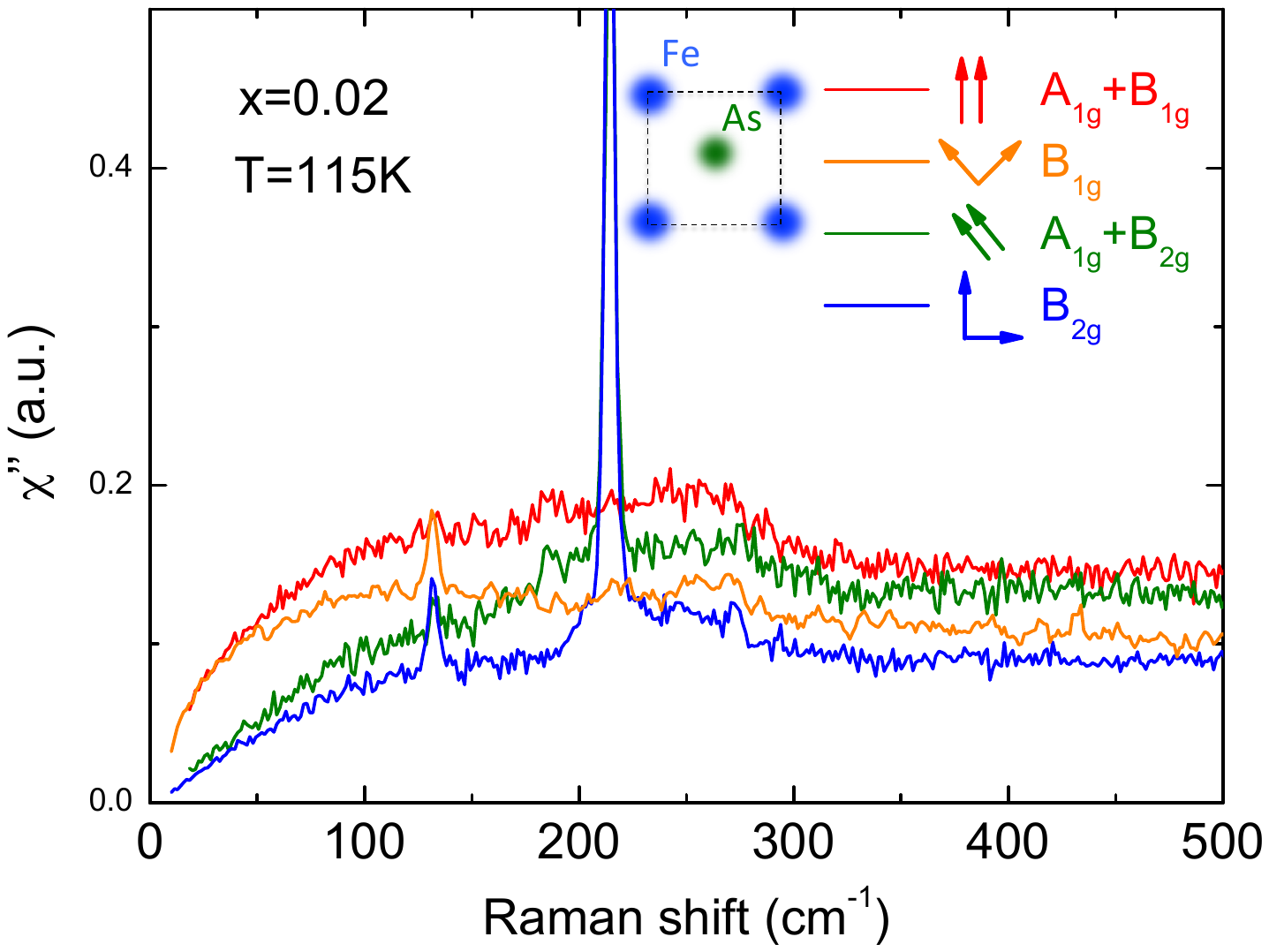}
\caption{Raman spectra in four polarization configurations for the $x=0.02$ composition (T$_{s}$=110K)
 and at $T=115K$. For each configuration, the polarizations of the incoming  and outgoing photons are sketched with respect to the Fe-As plane. The symmetries indicated correspond to the 1 Fe unit cell.}
\label{polar} 
\end{figure}

\section{Raman Vertex in the orbital basis}

In a multi-orbital system, the basic operator for Raman response in
the symmetry channel $\mu$ is given by: 
\begin{equation}
\rho^{\mu}=\sum_{\mathbf{k}\sigma}\sum_{m,n}\gamma_{mn}^{\mu}\left(\mathbf{k}\right)c_{m,\mathbf{k}\sigma}^{\dagger}c_{n,\mathbf{k}\sigma}^{\phantom{\dagger}}\label{charge_eff}
\end{equation}
 where $c_{m,\mathbf{k}\sigma}^{\dagger}$ creates an electron with
momentum $\mathbf{k}$ and spin $\sigma$ in orbital $m$. For the
$x^{2}-y^{2}$ channel, ignoring screening effects and vertex corrections,
the free-electron form factor or Raman vertex depends only on the
band dispersion $\varepsilon_{mn}\left(\mathbf{k}\right)$:

\begin{equation}
\gamma_{mn}^{x^{2}-y^{2}}\left(\mathbf{k}\right)=\frac{\partial^{2}\varepsilon_{mn}\left(\mathbf{k}\right)}{\partial k_{x}^{2}}-\frac{\partial^{2}\varepsilon_{mn}\left(\mathbf{k}\right)}{\partial k_{y}^{2}}\label{vertex_eff}
\end{equation}

In the iron pnictides, all $3d$ orbitals may contribute to the Fermi
surface. Let us focus on the role played by the orbitals $xz$ and
$yz$. In the tetragonal phase, their intra-orbital dispersions are
identical upon a $90^{\circ}$ rotation of the coordinate system,
i.e.

\begin{equation}
\varepsilon_{xz,xz}\left(k_{x},k_{y}\right)=\varepsilon_{yz,yz}\left(-k_{y},k_{x}\right)\label{eq1}
\end{equation}

Therefore, it follows for the intra-orbital Raman vertices:

\begin{equation}
\gamma_{xz,xz}^{x^{2}-y^{2}}\left(k_{x},k_{y}\right)=-\gamma_{yz,yz}^{x^{2}-y^{2}}\left(-k_{y},k_{x}\right)\label{eq2}
\end{equation}

\begin{figure}
\centering \includegraphics[clip,width=0.95\linewidth]{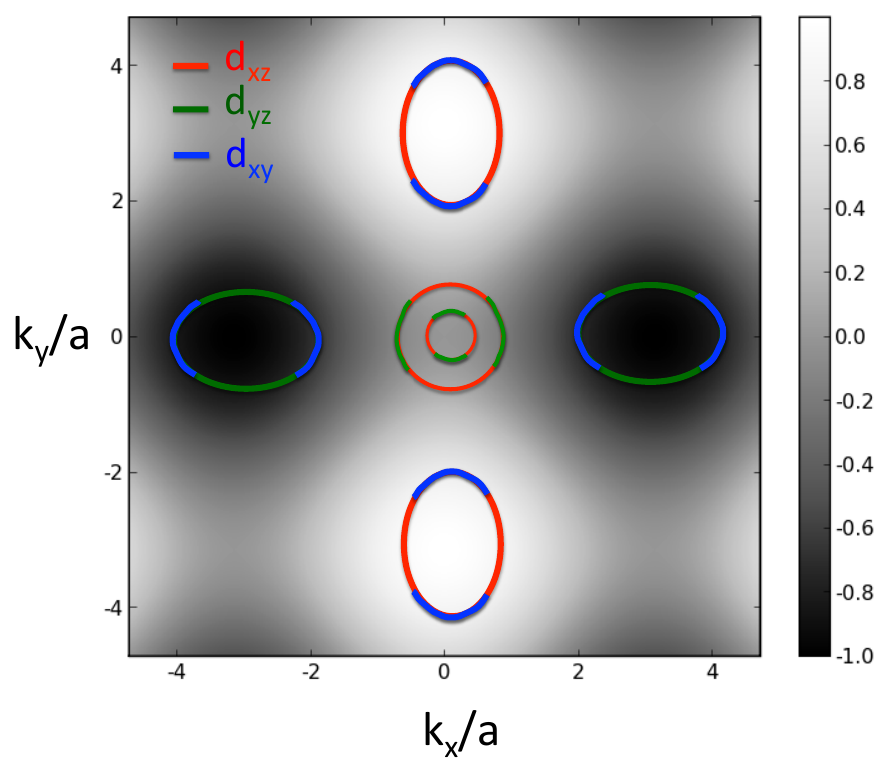}
\caption{$x^{2}-y^{2}$ form factor ($\cos k_{x}-\cos k_{y}$) in grey scale.
A sketch of the typical Fermi surface sheets of the iron pnictides
has been superimposed \cite{Graser}. Their orbital content is also
indicated.}
\label{fig2} 
\end{figure}

At low energies, the main contribution to the Raman scattering comes
from electronic states at the Fermi level. Symmetry requires that
if a point $\left(k_{x},k_{y}\right)$ at the Fermi surface has $xz$
orbital character, then the point at $\left(-k_{y},k_{x}\right)$
also belongs to the Fermi surface and has $yz$ orbital character.
Combined with the symmetry relation (\ref{eq2}), this implies that
the charge density difference $\delta n\equiv n_{xz}-n_{yz}$, with
the proper overall form factor, contributes to the effective $x^{2}-y^{2}$
Raman charge.

Similar conclusions can be drawn by analyzing the orbital content
of the Fermi surface of the iron pnictides and the $x^{2}-y^{2}$
form factor $\cos k_{x}-\cos k_{y}$. In Fig. \ref{fig2}, we superimpose
a sketch of the typical Fermi surface of the iron pnictides to the
form factor plotted in Fig. 1(c) of the main text. It is clear that
the form factor changes sign between the electron pockets located
at $X=\left(\pi,0\right)$ and $Y=\left(0,\pi\right)$. First principle
calculations \cite{Graser} reveal that while the $X$ electron pocket
has mostly $d_{yz}$ character (and no $d_{xz}$ contribution), the
$Y$ pocket has mostly $d_{xz}$ character (and no $d_{yz}$ contribution).
Therefore, since the $x^{2}-y^{2}$ form factor changes sign between
the $X$ and $Y$ pockets, and they have symmetry-related $d_{xz}$
and $d_{yz}$ spectral weights, the relative charge $\delta n$ between
the $d_{xz}$ and $d_{yz}$ orbitals appears in the $x^{2}-y^{2}$
response.

Notice that the remaining $3d$ orbitals, as well as inter-orbital
terms, in principle, also contribute to the $x^{2}-y^{2}$ Raman response.
A detailed discussion of all the non-zero intra-orbital and inter-orbital
Raman vertices $\gamma_{mn}^{\mu}\left(\mathbf{k}\right)$ was presented
in Ref. \cite{Valenzuela}. For instance, the $xy$ intra-orbital
vertex also satisfies the relationship:

\begin{equation}
\gamma_{xy,xy}^{x^{2}-y^{2}}\left(k_{x},k_{y}\right)=-\gamma_{xy,xy}^{x^{2}-y^{2}}\left(-k_{y},k_{x}\right)\label{eq2b}
\end{equation}
 implying that a breaking of tetragonal symmetry could in principle
be driven solely by the $xy$ orbital. However, if the nematic instability
is driven by a spontaneous $d_{xz}/d_{yz}$ orbital polarization,
as suggested by a few theoretical models, then the fluctuations associated
with the charge density difference $\delta n$ between $xz$ and $yz$
orbitals $\left\langle \delta n\,\delta n\right\rangle $ should dominate
the $x^{2}-y^{2}$ Raman susceptibility.

\end{document}